\documentclass[12pt, preprint]{aastex}
\usepackage{amsmath}
\usepackage{graphicx}
\newcommand{\kerrbbtwo}{{\sc kerrbb2~}}

\newcommand{\simplb}{{\sc simpl}}

\newcommand{\bhspecb}{{\sc bhspec}}

\newcommand{\Mdot}{\dot{M}}

\newcommand{\msun}{\rm M_{\sun}}

\newcommand{\rchinu}{\chi^{2}/\nu}
\newcommand{\fsc}{f_{\rm SC}}
\newcommand{\nh}{N_{\rm H}}

\newcommand{\kpc}{\rm kpc}

\newcommand{\keV}{\rm keV}

\newcommand{\cm}{\rm cm}

\newcommand{\rin}{r_{\rm in}}
\newcommand{\Rin}{R_{\rm in}}
\newcommand{\risco}{R_{\rm ISCO}}

\newcommand{\ledd}{L_{\rm Edd}}
\newcommand{\lledd}{l_{\rm D}} %L_{\rm D}/L_{\rm Edd}}

\newcommand{\rxte}{{\it RXTE~}}

\newcommand{\xmm}{{\it XMM~}}

\newcommand{\swift}{{\it Swift~}}
\newcommand{\bepposax}{{\it BeppoSAX~}}
\newcommand{\suzaku}{{\it Suzaku~}}
\newcommand{\ginga}{{\it Ginga~}}
\newcommand{\asca}{{\it ASCA~}}
\newcommand{\exosat}{{\it EXOSAT~}}

\newcommand{\rxteb}{{\it RXTE}}

\newcommand{\xmmb}{{\it XMM}}
\newcommand{\xmmnewtonb}{{\it XMM-Newton}}
\newcommand{\swiftb}{{\it Swift}}
\newcommand{\bepposaxb}{{\it BeppoSAX}}
\newcommand{\suzakub}{{\it Suzaku}}
\newcommand{\gingab}{{\it Ginga}}
\newcommand{\ascab}{{\it ASCA}}
\newcommand{\exosatb}{{\it EXOSAT}}

\shorttitle{ LMC X--3: A Basis for Measuring BH Spin  }
\shortauthors{Steiner et al.}

\begin{document}

\title{The Constant Inner-Disk Radius of LMC X--3: A Basis for Measuring Black Hole Spin}

\author{James F.\ Steiner\altaffilmark{1}, Jeffrey
  E.\ McClintock\altaffilmark{1}, Ronald
  A.\ Remillard\altaffilmark{2}, Lijun Gou\altaffilmark{1}, Shin'ya
  Yamada\altaffilmark{3}, Ramesh Narayan\altaffilmark{1}}

\altaffiltext{1}{Harvard-Smithsonian Center for Astrophysics, 60
  Garden Street, Cambridge, MA 02138.} \altaffiltext{2}{MIT Kavli
  Institute for Astrophysics and Space Research, MIT, 70 Vassar
  Street, Cambridge, MA 02139.}  \altaffiltext{3}{Department of
  Physics, University of Tokyo, 7-3-1, Hongo, Bunkyo-ku, Tokyo,
  113-0033}

\email{jsteiner@cfa.harvard.edu}

\begin{abstract}

The black-hole binary system LMC X--3 has been observed by virtually
every X-ray mission since the inception of X-ray astronomy.  Among the
persistent sources, LMC X--3 is uniquely both habitually soft and
highly variable.  Using a fully relativistic accretion-disk model, we
analyze hundreds of spectra collected during eight X-ray missions that
span 26 years.  For a selected sample of 391 {\it RXTE} spectra we
find that to within $\approx 2$~percent the inner radius of the accretion
disk is constant over time and unaffected by source variability.  Even
considering an ensemble of eight X-ray missions, we find consistent
values of the radius to within $\approx 4-6$~percent.  Our results
provide strong evidence for the existence of a fixed inner-disk
radius.  The only reasonable inference is that this radius is closely
associated with the general relativistic innermost stable circular
orbit (ISCO).  Our findings establish a firm foundation for the
measurement of black hole spin.

\end{abstract}


\section{Introduction}\label{section:Intro}

The X-ray binary LMC X--3 was discovered by {\it Uhuru} in 1971
(\citealt{Leong_1971}).  Observations of its B3V optical counterpart
revealed an orbital period of 1.7~days and a mass function of
$2.3\pm0.3~\msun$.  Because of its massive companion star, this
established LMC X--3 as a strong dynamical black-hole (BH) candidate
\citep{Cowley_1983, Kuiper_1988}.  Subsequent X-ray observations
spanning decades have revealed a complex behavioral pattern that
includes transitions between soft and hard states \citep{Wilms_2001}
and long-term ($\gtrsim100$~d) variability cycles \citep{Cowley_1991}.
While by some metrics LMC X--3 is a nearly archetypal BH binary, its
combined qualities of persistence and strong variability set it apart
as unique.

Among the black hole systems, LMC X--3 bridges the divide between
low-mass X-ray binaries powered by Roche-lobe overflow and wind-fed,
high-mass X-ray binaries \citep{Soria_2001}.  The former are
transients, usually locked in a deep quiescent state, whereas the
latter systems are persistently X-ray bright.  Among the classical
persistent BH sources (Cyg X--1, LMC X--1, and LMC X--3), LMC X--3
habitually shows the softest X-ray spectrum, reaches the highest
luminosity, and exhibits the largest variations in intensity.

Because of its persistence LMC X--3 has been observed by nearly every
X-ray astronomy mission.  In this Letter, we apply our relativistic
accretion disk model ({\sc kerrbb2}; \citealt{McClintock_2006}) to
essentially all available X-ray data in order to examine the presumed
constancy of the inner radius of the BH's accretion disk.  We draw
upon data collected by eight missions, with {\it RXTE} providing the
lion's share.

For thin accretion disks, recent MHD simulations provide support for
identifying the inner-disk radius $\Rin$ with the radius of the
innermost stable circular orbit $\risco$
(\citealt{Reynolds_Fabian_2008, Shafee_2008, Penna_2010}; but see
\citealt{Noble_2009}), a proposition that has a long history of
theoretical and observational support (e.g., see Section 6 in
\citealt{Gou_2009}).  With this identification and the simple
monotonic relationship between $\risco$ and the BH spin parameter
\citep{Shapiro_Teukolsky}, a measurement of $\Rin$ is equivalent to a
measurement of the spin of the BH.  This is the basis for both the
continuum-fitting \citep{Zhang_1997} and Fe-K \citep{Fabian_1989}
methods of measuring spin.  In recent years, both methods have been
used to estimate the spins of stellar BHs (e.g., \citealt{Shafee_2006,
  McClintock_2006, Davis_2006, Reis_2008, Miller_2009, Blum_2009}).

The mass of LMC X--3's BH primary is presently very uncertain
\citep{Cowley_1992}, and we adopt a round value that is typical for BH
binaries of $M=10~\msun$.  For the inclination we adopt the
provisional value $i=67\degr$ \citep{Kuiper_1988}.  Because of the
uncertainties in both $M$ and $i$, in this Letter we do not attempt to
estimate the BH's spin.  Rather, we assume a reasonable value for the
mass and employ the X-ray continuum-fitting method in order to study
the constancy of $\Rin$.  We describe our data set comprised of
hundreds of multi-mission spectra in Section
\ref{section:Observations}, our analysis in Section
\ref{section:Analysis}, and present our results in Section
\ref{section:Results}.  In Section \ref{section:Discussion}, we
explore the systematics associated with our spectral model and
conclude by discussing our results in the context of ongoing studies
of BH spin.

%%%%%%%%%%%%%%%%%%%%%%%%%%%%%%%%%%%%%%%%%%%%%%%%%%%%%%%%%%%%%%%%%%%%%%%%%%

\section{Observations}\label{section:Observations}

{\it \rxteb}: The {\it Rossi X-ray Timing Explorer} (\rxteb) is our
workhorse instrument, providing a total gross sample of 712 spectra.
Individual spectra were defined by grouping all the archival pointed
data from 1996 through 2009 into approximately half-day bins with
$\approx 90\%$ of exposure times ranging from 1--10~ks.  We only use
pulse-height spectra obtained by the best-calibrated PCA detector,
PCU-2 \citep{Jahoda_2006}.  Count rates have been renormalized to
correct for detector dead time and a systematic error of 1\% has been
included to account for uncertainty in the instrumental response
\citep{Jahoda_2006}.  These data have been analyzed from 2.55--25 keV
over all reliable gain epochs ($\geq$ epoch 2).  Here and elsewhere,
the analysis work has been performed using XSPEC v.12.5.1o (recent
enough that an early coding error in kerrbb has been
fixed\footnote{http://heasarc.nasa.gov/docs/xanadu/xspec/issues/archive/issues.12.5.0an.html};
\citealt{XSPEC}).

{\it \exosatb}: Seven observations from 1983--1984 were obtained via
the HEASARC archive\footnote{http://heasarc.nasa.gov}; only data from
the ME instrument are currently available.  Spectra were extracted as
described in \citet{Treves_Belloni_1988} and analyzed from 1--25 keV.
The customary systematic error of 1\% was included.

{\it \gingab}: The LAC detector observed LMC X--3 on 18 occasions
during 1987--1990. To extract these spectra, we followed the
procedures described in \citet{Ebisawa_1993}.  Each spectrum has been
analyzed from 1.5-25 keV with a 1\% systematic error included.

{\ascab}: LMC X--3 was observed twice, once on UT 1993 September 22
and later on UT 1995 April 14.  We extracted and separately combined
spectra from the two GIS and two SIS instruments.  Data were
calibrated relative to the GIS-2 detector and analyzed from 0.8-9~\keV
(GIS) and 0.6-9~\keV (SIS) using a 2\% systematic uncertainty.

{\it \bepposaxb}: Following the standard reduction guide
(\citealt{BepposaxABC}), we have generated spectra for the
narrow-field instruments from each of the 23 available observations.
We extracted spectra using 8$\arcmin$ apertures in the imaging
instruments and used a fixed rise-time threshold for the PDS.  For
each observation we employed all usable LECS, MECS, and PDS data.
Throughout, we adopted the standard inter-detector floating
normalizations calibrated relative to the MECS.  Data were analyzed
from 0.12--4 keV (LECS), 1.65--10 keV (MECS) and 15--80 keV (PDS).  A
1\% systematic error has been included.

%1\% (Montonari et al. )

{\it \xmmnewtonb}: All photon-counting data were severely piled up and
therefore rejected because of uncertainties in the flux calibration.
We use the single available 19~ks timing-mode observation of LMC X--3
obtained on UTC 2000 November 25.  Because of the large number of
accumulated counts, $\sim 2\times10^6$, uncertainties in the response
of the detector are dominant, and we therefore included a 3\%
systematic error and fitted over $0.5-10~\keV$.  Reduction and
processing has been performed using \xmm SAS
v9.0.0\footnote{http://xmm.esac.esa.int/sas/}.

{\it \swiftb}: The sole XRT windowed-timing mode observation of LMC
X--3, taken on UTC 2007 November 26, has been procured and analyzed
following the procedures outlined in \citet{SWIFT:redux}.  
We rejected all the photon-counting data
because they suffer from extreme pile-up.  Calibration version 11
files have been used for the data reduction.  In consultation with the
\swift Help Desk, we have included an extra model component to account
for an instrumental artifact near the Si edge around 1.7~keV.
Analysis has been conducted over 0.4--10~keV using a 1\% systematic
error.

{\it \suzakub}: Two observations were made on Dec. 22 2008 and Dec. 21
2009 (UT).  The Suzaku attitude calibration was improved using the
{\it AEattcor}
routine\footnote{http://space.mit.edu/CXC/software/suzaku/}.  We
applied the appropriate reduction procedures for a bright point
source\footnote{http://www.astro.isas.ac.jp/suzaku/analysis/xis/pileup/HowToCheckPileup\_v1.pdf}.
Pile-up was kept well below $\sim 3\%$ by excluding the innermost
10$\arcsec$ and 30$\arcsec$ for the 2008 and 2009 observations,
respectively.  In all other respects, we have followed the methods of
\citet{Kubota_2010}, including using their energy intervals and
adopting a 1\% systematic error.  A fixed cross-normalization of 1.16
is used between XIS and HXD-PIN detectors \citep{Suzaku_TD}.

% The ancillary responses were refined using an exposure map to
% optimize the flux calibration.

\subsection{Flux Calibration}\label{subsection:flux}

Just as deducing the radius of a star from its spectrum requires
knowledge of its luminosity, in order to estimate the inner radius of
an accretion disk it is also necessary to determine its luminosity.
However, the measurement of X-ray luminosity is problematic in X-ray
astronomy because of the significant flux-normalization differences,
often $\gtrsim$10\%, between missions.  We address this issue by using
the power-law spectrum of the Crab Nebula as measured by
\citet*{Toor_Seward}: $\Gamma = 2.1$ and
$N=9.7$~photons~s$^{-1}$~keV$^{-1}$ at 1~\keV.

%We consider this measurement to still
% be the most reliable flux standard to date (but see also
% \citealt{Kirsch_2005}).

For each mission considered herein (excepting \swiftb; see Table~1),
we either rely on the Crab calibration performed by the instrument
team, or we compute a correction to the effective area by comparing
the spectrum predicted by \citet*{Toor_Seward} to parameters obtained
by analyzing proximate, archival observations of the Crab.  Toor \&
Seward normalization coefficients $f_{\rm TS}$ and slope differences
$\Delta\Gamma_{\rm TS}$ are presented for each mission in Table~1.
This table also summarizes for LMC X--3 the gross number of
observations available from each mission, $N_{\rm obs}$, as well as
the number of observations that meet our selection criteria, $N_{\rm
  sel}$ (Section \ref{subsec:selection}).

\section{Analysis}\label{section:Analysis}

At energies above $\sim 5-10$~keV, the spectra of BH binaries in all
states show a contribution from a power-law component.  This power law
is widely attributed to inverse-Compton scattering of thermal disk
photons by hot coronal electrons.  The power-law model we employ, {\sc
  simpl}, generates this Compton component by upscattering seed
photons from the thermal component (\citealt{Steiner_simpl}).

The thermal and principal component of our model is {\sc kerrbb2}, a
thin accretion disk model that includes all relativistic effects,
self-irradiation of the disk (``returning radiation''), limb
darkening, and the effects of spectral hardening \citep{KERRBB,
  McClintock_2006}.  During analysis, this latter effect is handled on
the fly via a look-up table of the spectral hardening factor $f$ for
a given value of the disk viscosity parameter $\alpha$ (we adopt
$\alpha=0.01$ as default).  These tables were computed using \bhspecb,
a second relativistic disk model \citep{Davis_2006, BHSPEC}.

Our fit to the thermal component of the spectrum effectively
determines the solid angle subtended by the accretion disk: $\Omega =
{\pi (R_{\rm in}/D)^2}$cos~$i$, where $D$ is the distance and $i$ is
the inclination of the accretion disk with respect to the line of
sight.  For $D$ we use the average distance to the LMC, $D=48.1~\kpc$
(e.g., \citealt{Orosz_LMCX1}), while for inclination we use
$i=67\degr$ (Section 1).  Finally, we express $\Rin$ in dimensionless
form, $\rin \equiv \Rin/{(GM/c^2)}$ using $M=10~\msun$~(Section 1).
We have recently shown that the choice of $M$, $i$ and $D$, which
effectively sets the absolute scale for $\rin$ and the luminosity, is
quite unimportant for testing the stability of $\rin$ (see Fig.\ 3 and
text in \citealt{Steiner_2009}).  (These values are crucial, however,
when it comes to estimating the spin of the black hole.)

Using our adopted values of the source $M$, $i$, and $D$, our source
model has four fit parameters: two for {\sc kerrbb2}, $\Rin$ and the
mass accretion rate $\Mdot$, and two for {\sc simpl}, the photon index
$\Gamma$ and $\fsc$, which is the fraction of disk photons that get
re-directed via scattering into the power law.  Our full model is {\sc
  tbabs(simpl$\otimes$kerrbb2)}, where {\sc tbabs} models the effects
of photoelectric absorption; we fix its sole parameter:
$\nh=4\times10^{20}\cm^{-2}$ \citep{Page_2003}, using abundances from
\citet{Wilms_2000}.  For \kerrbbtwo we include limb darkening and
returning radiation effects, set the torque at the inner boundary of
the accretion disk to zero, and fix the normalization to unity.  We
use the efficient, up-scattering-only version of \simplb, and in
Section \ref{section:Discussion} we show that this choice is
unimportant.

%%%%%%%%%%%%%%%%%%%%%%%%%%%%%%%%%%%%%%%%%%%%%%%%%%%%%%%%%%%%%%
  \begin{deluxetable}{lccrrrcc} 
%  \rotate 
  \tabletypesize{\scriptsize} 
  \tablecolumns{8}
  \tablewidth{0pc}  
  \tablecaption{Data and Instrument Summary}

  \tablehead{Instrument & $N_{\rm obs}$ & $N_{\rm sel}$\tablenotemark{a} & & $f_{\rm TS}$\tablenotemark{b} & $\Delta\Gamma _{\rm TS}$\tablenotemark{b}  & Ref. \\
} \startdata  

\rxte (PCU-2)    &  712  &  391~(568) &&     1.097  &     0.010 & \nodata \\
\tableline
\suzaku (XIS0)   &    2  &    2~(2) &&     0.98  &    -0.01 &   1,2 \\
\swift (XRT)     &    1  &    1~(1) &&     1.10  &    -0.04 &   3\tablenotemark{c} \\
\xmm (MOS-1)     &    1  &    0~(1) &&     1.00  &     0.01 &   4,5,6  \\ 
\bepposax (MECS) &   23  &   2~(23) &&     0.95  &     0.00 &   7\\
\asca (GIS-2)    &    2  &    2~(2) &&     0.97  &    -0.01 &   8,9\\
\ginga  (LAC)    &   18  &   7~(11) &&     0.94  &    -0.02 &   10\\
\exosat (ME)     &    7  &    6~(6) &&     0.98  &     0.00 &   11\\

\enddata

\tablerefs{ (1) \citealt{Suzaku_XIS}; (2) {http://heasarc.gsfc.nasa.gov/docs/suzaku/prop\_tools/suzaku\_td/suzaku\_td.html}; (3)
  \citealt{Godet_2009}; (4) \citealt{XMM:CAL-TN-0083}; (5)
  \citealt{XMM:CAL-TN-0052}; (6) private communication with Ignacio de
  la Calle; (7) \citealt{BepposaxABC}; (8) \citealt{Makishima_1996};
  (9) \citealt{ASCA:cal}; (10) \citealt{Turner_1989}; (11)
  \citealt{EXOSAT:cal} }

\tablenotetext{a}{ Number of selected observations.  Parentheses
  indicate the selection numbers when high luminosities $\lledd>0.3$
  are allowed (see Fig.~\ref{f2}).}

\tablenotetext{b}{ $f_{\rm TS}$ is the ratio of the Crab normalization
  to that of Toor \& Seward and $\Delta\Gamma _{\rm TS}$ is the
  difference between photon indices.}

\tablenotetext{c}{The \swift values are derived from a comparison
  between \rxte and \swift observations of 3C 273.}

%EXOSAT: http://heasarc.gsfc.nasa.gov/docs/exosat/express/pg194.html#pg194

% ASCA: http://heasarc.gsfc.nasa.gov/docs/asca/xrt_new_response_announce/announce.html

\label{tab:obs}
\end{deluxetable}

%%%%%%%%%%%%%%%%%%%%%%%%%%%%%%%%%%%%%%%%%%%%%%%%%%%%%%%%%%

\subsection{Data Selection}\label{subsec:selection}

Our preliminary analysis of all the data showed that for many spectra
the power-law index $\Gamma$ was essentially unconstrained, even for
the \bepposaxb, \exosatb, \gingab, and \rxte missions, which have the
requisite coverage to detect this component.  This is because the
source is relatively faint ($\lesssim 50$~mCrab) and its Compton
power-law component is generally very weak, showing a median
normalization $\fsc \approx 0.3\%$.  The extreme dominance of the thermal
component in LMC X--3 makes it an ideal source for accretion-disk
studies such as this.

Restricting our census to the 134 \rxte spectra for which the photon
index is measured to a precision better than $\sigma_\Gamma=0.5$, we
find a strong clustering of values in the range $\Gamma \approx
2-2.6$.  For our baseline model we fix $\Gamma=2.35$ which matches the
constant index derived from 22 deep \rxte pointings by
\citet{Smith_2007}, and in Section \ref{section:Discussion} we show
that our results depend very weakly on this choice for
$2\lesssim\Gamma\lesssim3$.  

Meanwhile, 
three missions, \ascab, \swift and \xmmb, have no sensitivity above
$E\approx$10~keV, and therefore only very loosely measure the
power-law normalization parameter, $\fsc$.  At the same time, a self
consistent and fruitful analysis of the thermal and Compton components
requires that $\fsc$ be sensibly constrained.  Therefore, and because
the power law is generally so weak, we impose an additional
data-selection requirement, namely that for each fit $\fsc$ falls
within the lower 95\% span of the \rxte rank-ordered values.

We further adopt a goodness-of-fit requirement, $\rchinu < 2$, and a
lower limit on the Eddington-scaled disk luminosity, $\lledd \equiv
L_{\rm D}/\ledd > 0.05$.  This latter criterion removes any hard state
data in which the disk is likely truncated at $r > \rin$ (e.g.,
\citealt{Esin_1997}).  Finally, in consonance with the thin-disk model
employed, we select only data for which $\lledd < 0.3$
\citep{McClintock_2006}.

\begin{figure} 
\plotone{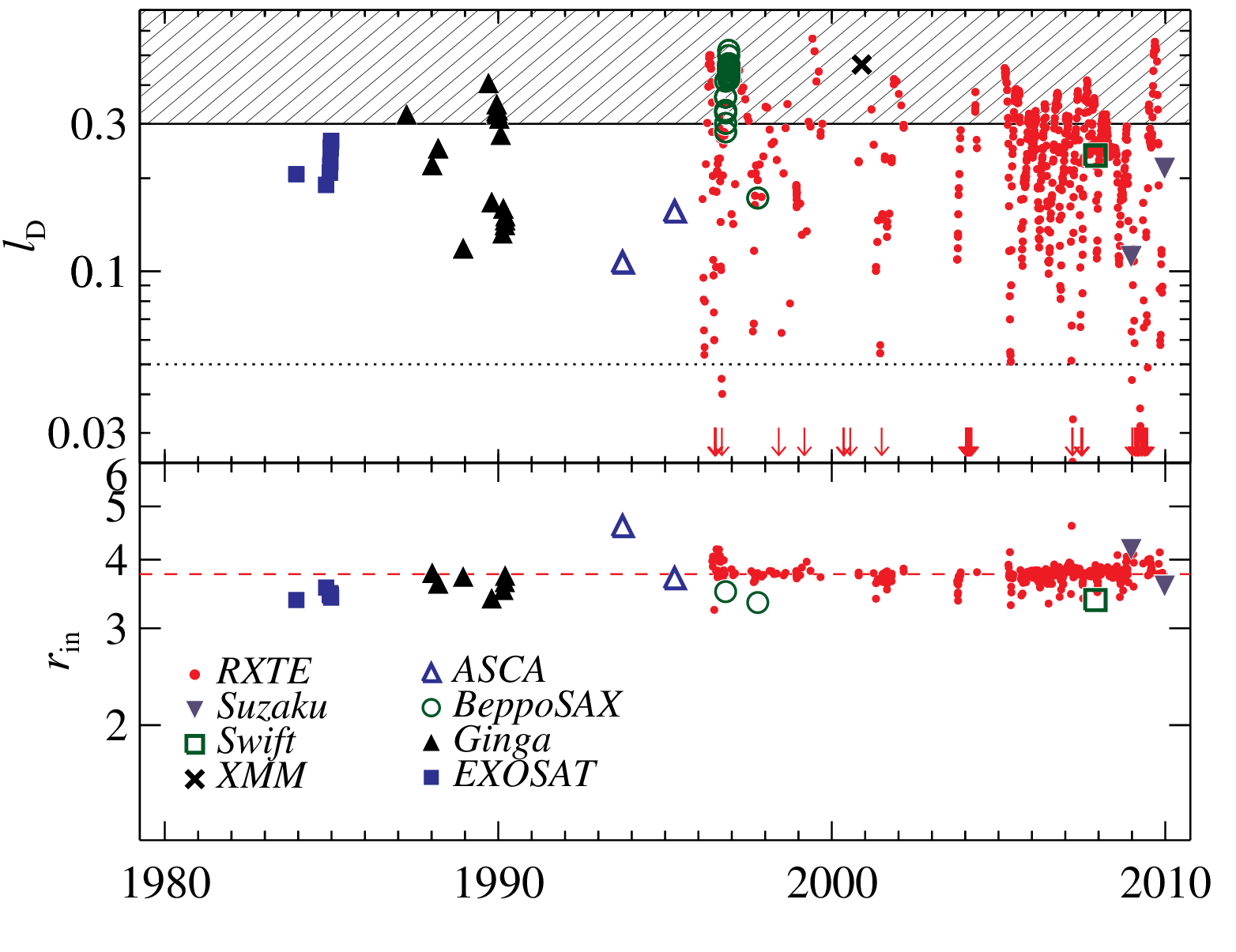} 
\caption{ {\bf { top:}} Accretion-disk luminosity in Eddington-scaled
  units ($M=10~\msun$) versus time for all the data considered in this
  study (766 spectra).  Red arrows show \rxte data which are off
  scale.  Data in the unshaded region satisfy our thin-disk selection
  criterion ($H/R < 0.1$, which implies $ \lledd < 0.3$;
  \citealt{McClintock_2006}).  The dotted line indicates the lower
  luminosity threshold (5\% $\ledd$) adopted in Section
  \ref{subsec:selection}.  {\bf { bottom:}} Values of the
  dimensionless inner disk radius $\rin$ are shown for thin-disk data
  in the top panel that meet all of our selection criteria (411
  spectra; see Section \ref{subsec:selection}).  Despite large
  variations in luminosity, $\rin$ remains constant to within $\approx
  4\%$ over time.  The median value for the \rxte data alone ($\rin =
  3.77$) is shown as a red dashed line.  }

\label{f1} 
\end{figure}

\section{Results}\label{section:Results}

The top panel of Figure~\ref{f1} shows a 26-year record of the disk
luminosity of LMC X--3, which is seen to vary by orders of magnitude.
Two-thirds of the data meet our thin-disk selection criterion $\lledd
< 0.3$.  In the lower panel, we show the time history of the inner
disk radius $\rin$ for just those data that meet all of our selection
criteria (Section \ref{subsec:selection}).  The radius is constant
over the 26 years of monitoring to within $\sim 2\%$ for \rxte alone
and $\sim 4\%$ considering all missions.

%The sole \xmm observation is too luminous, and so
%only 7 of the 8 missions appear in this panel.

\begin{figure} 
\plotone{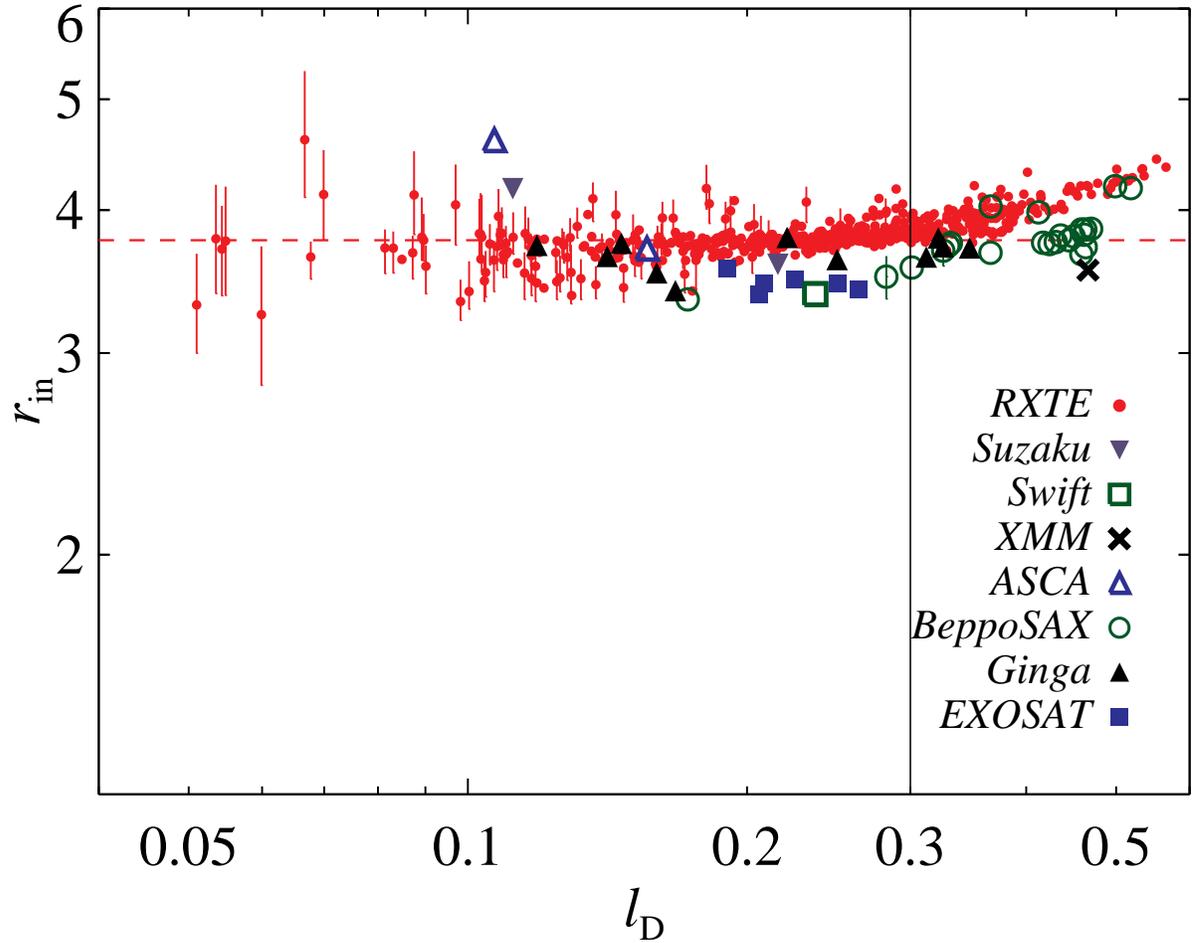}
\caption{The dimensionless inner-disk radius $\rin$ versus luminosity
  for the filtered data (Section \ref{subsec:selection}) and our
  baseline model.  The vertical black line shows our adopted thin-disk
  upper limit, $\lledd = 0.3$.  As in Figure~\ref{f1}, the red-dashed
  line shows the \rxte average below this limit. }
\label{f2} 
\end{figure}

Figure~\ref{f2} explores the dependence of $\rin$ on luminosity.  In
this figure we include the high-luminosity data ($\lledd > 0.3$) that
meet all of our other selection criteria (Section
\ref{subsec:selection}).  For $\lledd < 0.3$ there is a gentle,
nonlinear rise of $\rin$ with luminosity.  Especially visible in the
\rxte data, this rise becomes prominent beyond $\lledd \sim 0.25$,
above which there is a $\sim 12\%$ increase in $\rin$.  No significant
change in $\rchinu$ is associated with the apparent increase of
$\rin$.  We cannot say if this represents a real increase in $\rin$ at
high luminosities or is simply an artifact of using the thin-disk
model, which is expected to be increasingly inaccurate at higher
luminosities \citep{Penna_2010, Abramowicz_2010} at which a transition
may occur to an advective slim-disk accretion mode.  Interestingly,
however, despite this rapid rise, we note that the \rxte data appear
tightly clustered along a well-defined curve.  We approximate this
dependence using a non-parametric curve-fit (LOWESS; \citealt{LOWESS})
that allows us to detrend the data.  We conclude that results from all
eight missions, including the high-luminosity data, are in agreement
with one another to within $\approx 6\%$.  

%This is a few times the internal variability of 2\% measured from the
%\rxte data alone.

\section{Discussion}\label{section:Discussion}

\begin{figure} 
\plotone{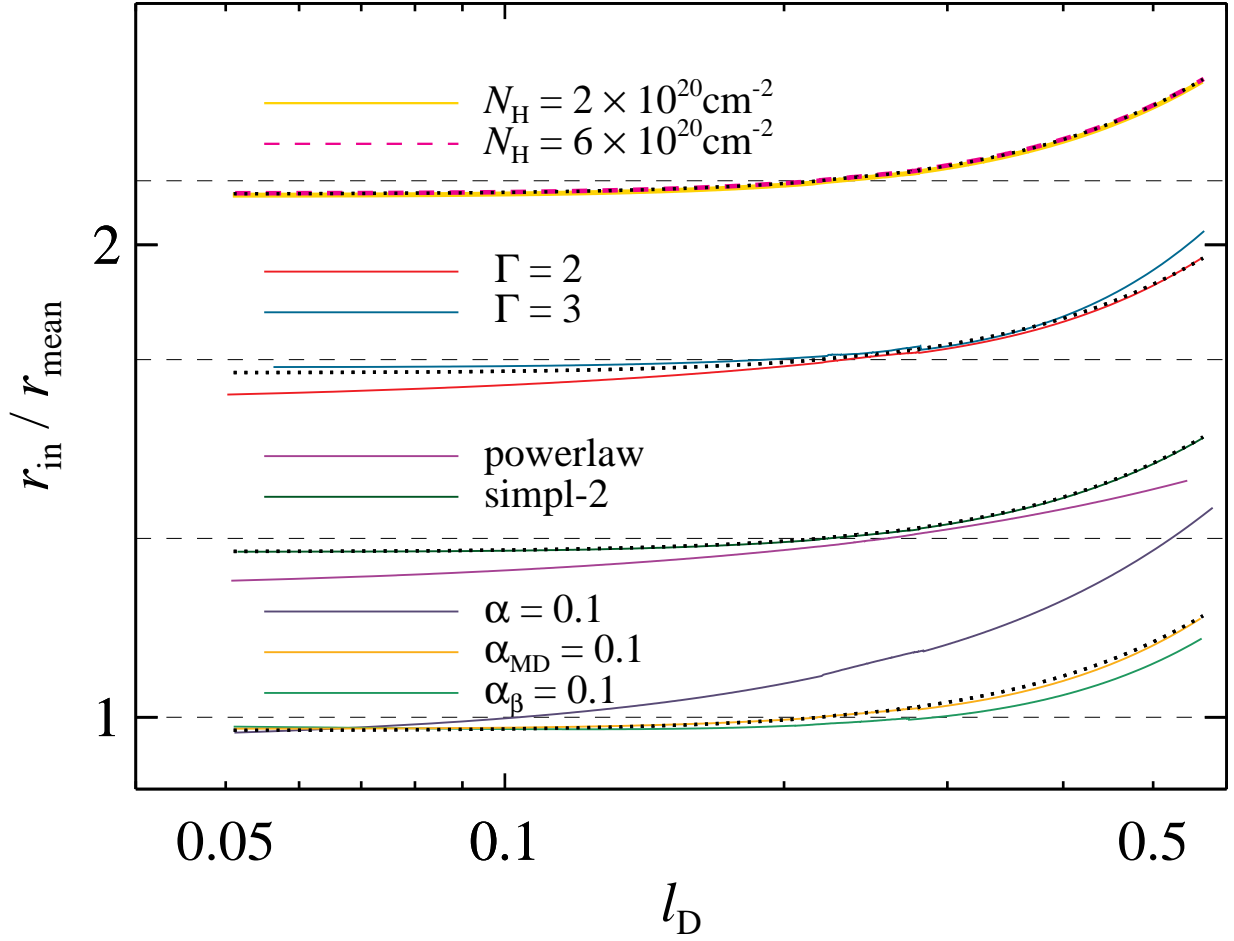} 
\caption
{ Four families of models showing how our baseline results in
  Figure~\ref{f2} are affected when a single model component or
  parameter is varied.  The black dotted line drawn with each family
  of curves represents our fiducial model: $\nh=4 \times 10^{20}
  \cm^{-2}$; $\Gamma=2.35$; {\sc simpl-1}; and $\alpha=0.01$.  The
  horizontal dashed line for each family is set by the average value
  of $\rin$ (see Figures~\ref{f1}~\&~\ref{f2}), and each family is
  offset by 30\% for clarity.  Each curve represents a LOWESS curve
  fit to the \rxte data alone.  Both axes are scaled logarithmically.
}
\label{f3}
\end{figure}

Figure~\ref{f2} clearly demonstrates the limitations of the thin-disk
model at high luminosities.  We further illustrate this point in
Figure~\ref{f3} using LOWESS fits to the abundant \rxte data.  We
vary, one-at-a-time, the model components and parameters of our
baseline model, grouping these trials into four separate ``families.''
In order of increasing importance, these families are (1) column
density $\nh$, (2) power-law index $\Gamma$, (3) choice of power-law
model, and (4) $\alpha$.  Figure~\ref{f3} illustrates the changes
introduced by adjusting each family of settings.

We highlight two conclusions from Figure~\ref{f3}: (1) Our results are
relatively insensitive to all settings with the single exception of
the choice of $\alpha$-viscosity; the value $\alpha=0.1$ increases
significantly the dependence of $\rin$ on luminosity.  (2) The
positive correlation between $\rin$ and luminosity is generally
present for all families over the full range of luminosity, but it
becomes prominent only above $\lledd \approx 0.2-0.3$.

Inspecting the families of curves in Figure 3 from top to bottom, one
concludes the following: As the first two families show, our results
are insensitive to the choice of $\nh$ and only modestly sensitive to
the choice of $\Gamma$.  In modeling the Compton tail component (third
family), one sees that our results are essentially identical whether
one uses our baseline up-scattering-only model {\sc simpl} $\equiv$
{\sc simpl-1} or a a two-sided scattering model {\sc simpl-2}
\citep{Steiner_2009}, while the results obtained using the standard
power law model {\sc powerlaw} differ only modestly ($\lesssim 5$\%).

The fourth family considers the primary setting for {\sc bhspec}, the
viscosity parameter $\alpha$, used to compute spectral hardening
(Section \ref{section:Analysis}).  Here, we examine several distinct
cases: our fiducial value, $\alpha = 0.01$ (dotted), the value $\alpha
= 0.1$ (Section~\ref{section:Analysis}; dark blue), and alternative
stress prescriptions $\alpha_{\rm MD} = 0.1$ (orange) and
$\alpha_\beta = 0.1$ (green).  The parameter $\alpha$ typically refers
to viscosity in the disk which is proportional to the total pressure
at the disk midplane.  However, other choices exist such as ``beta
disk'' and ``mean disk'' models in which $\alpha_\beta$ and
$\alpha_{\rm MD}$ respectively describe viscosities which scale
proportionally to the gas pressure or the geometric mean of gas and
total pressures \citep{Done_Davis_2008}.  Both latter options produce
spectral hardening values quite similar to those obtained for
$\alpha=0.01$.  In conclusion, only the second option, $\alpha = 0.1$,
has an important effect on our results.

%We have also investigated the importance of the metallicity setting in
%\bhspec (not shown in the figure), and find that values from $Z = 0.1
%Z_\sun$ to our fiducial solar value are indistinguishable.  

Our results indicate that the value of the inner disk radius $\rin$ --
and hence spin -- is stable over decades, as is expected given the
minute effects of accretion torques on a BH over such a time scale.
We also confirm that $\rin$ is nearly independent of luminosity
provided that the disk is geometrically thin.  The stability of $\rin$
over time (for $\lledd < 0.3$) despite large fluctuations in the mass
accretion rate provides strong evidence that $\rin$ and $\risco$ are
closely associated, as we tacitly assume in measuring BH spin (Section
\ref{section:Intro}).

The inter-mission consistency of our results ($\approx 4\%$ below
$\lledd < 0.3$ and 6\% overall) is very important for future X-ray
continuum measurements of BH spin: For some transient BH sources
(e.g., A0620--00 and GRS~1009--45) only one or a few spectra are
available in the data archives.  Our results for LMC X--3 show that,
as long as the power-law component is reliably measured, even a
single, suitable spectrum can deliver an estimate of the disk inner
radius accurate to several percent, and thereby a reliable measurement
of spin.

\acknowledgements

This research has made use of data obtained from the High Energy
Astrophysics Science Archive Research Center (HEASARC), provided by
NASA's Goddard Space Flight Center.  We thank Shane Davis and Laura
Brenneman for their insightful analyses of the manuscript and
suggestions.  JFS thanks Tim Oosterbroek for assisting with \bepposax
LECS, and Ignacio de la Calle for advice on {\it XMM}.  The authors
thank Tomaso Belloni and Ken Ebisawa for contributing reduced archival
data from \exosatb, and \ginga and \asca respectively.  JFS was
supported by the Smithsonian Institution Endowment Funds and JEM
acknowledges support from NASA grant NNX08AJ55G.  RN acknowledges
support from NASA grant NNX08AH32G and NSF grant AST-0805832.

%\clearpage
\newcounter{BIBcounter}        % Make a counter to reference 
\refstepcounter{BIBcounter}

\end{document}